\documentclass[showpacs,prl,twocolumn,aps]{revtex4}

 \usepackage{bm}
 \usepackage{amsmath}
 \usepackage{amssymb}
 \usepackage{latexsym} 
 \usepackage{amsfonts} 
 \usepackage{epsfig}
 \usepackage{psfrag}

 \newcommand{ \ket}[1]     { \left|#1\right> } 
 \newcommand{ \bra}[1]     { \left<#1\right| } 
  
 \newcommand{ \nn }        { \nonumber\\ } 
  
 \newcommand{ \f }[1]      { \mbox{\boldmath$#1$} }

 \newcommand{ \na }        { \mbox{\boldmath$\nabla$} }
 \newcommand{ \bea }       { \begin{eqnarray} }
 \newcommand{ \ea }        { \end{eqnarray} }
 \newcommand{ \eea}        { \end{eqnarray} }
 \newcommand{ \wbea }      { \begin{widetext} \begin{eqnarray} }
 \newcommand{ \weea }      { \end{eqnarray} \end{widetext} }
 \newcommand{ \wea }       { \end{eqnarray} \end{widetext} }
 \newcommand{ \ord }       { {\cal O} }

 \newcommand{ \Hu }        { \mathfrak H }
 \newcommand{ \BF }[1]     { \boldsymbol{#1} }

\begin{document}

\title{Signatures of Planck-scale interactions in the cosmic microwave
  background?}   

\author{Friedemann Queisser, Michael Uhlmann, and Ralf Sch\"utzhold}

\affiliation{Institut f\"ur Theoretische Physik, 
Technische Universit\"at Dresden, D-01062 Dresden, Germany}

\begin{abstract}
Based on a rather general low-energy effective action
(interacting quantum fields in classical curved space-times),
we calculate potential signatures of new physics 
(such as quantum gravity) at ultra-high energies 
(presumably the Planck scale) in the anisotropies of the cosmic 
microwave background.
These Planck-scale interactions create non-Gaussian contributions, 
where special emphasis is laid on the three-point function as the most 
promising observable, which also allows the discrimination between 
models violating and those obeying Lorentz invariance. 
\end{abstract}
 

\maketitle

{\em Introduction}\quad
%
In our present standard model of cosmology (see, e.g., \cite{text}), 
the anisotropies observed in the cosmic microwave background 
(e.g., by the recent WMAP \cite{wmap} 
and the future PLANCK \cite{planck} mission) originate from quantum
fluctuations of the inflaton field, which were amplified and stretched
by the cosmic expansion during inflation (a very early epoch in the
evolution of our universe).
The expansion of our universe acts as a cosmic microscope, i.e., 
these observed modes can be traced back to extremely short length
scales and thus ultra-high energies 
(such as the Planck energy $M_{\rm Pl}\approx10^{19}\,{\rm GeV}$) 
which are completely inaccessible with present or future terrestrial  
experiments (e.g., particle accelerators).
This observation entices the question of whether one might observe
signatures of new interactions (such as quantum gravity) beyond the
standard model of particle physics in the anisotropies.

Several different approaches for such expected deviations from our
known laws of physics at highest energies have been investigated
already.
One possibility to change the physics at small distances is a 
modification of the dispersion relation (of the inflaton field) 
implying a breakdown of the Lorentz invariance 
(which would then be just a low-energy symmetry), 
see, e.g., \cite{disp}.
It was also shown that an explicit short-distance cutoff has a
potentially measurable effect on the spectrum \cite{cutoff}.
Possible back-reaction effects resulting from UV-modifications have
been considered in \cite{back} and a possible change of the
propagator for the inflaton while preserving the local Lorentz
invariance has been studied in \cite{propagator}. 
Several authors have investigated different effects that result from 
a modification of the commutation relations at high energies such as 
$[x_i,p_j]=i\hbar\delta_{ij}(1+\eta\f{p}^2)$, see, e.g., 
\cite{commutation}.
Finally, the impact of different choices of the initial vacuum state
have been considered in \cite{initial}, for example.

These references \cite{initial} already illustrate the importance of 
considering the correct initial state.
(One could obtain basically arbitrary results by changing the initial
state correspondingly.)
Furthermore, the aforementioned investigations were mainly devoted to 
{\em linear} fields, whereas one would expect new physics such as
quantum gravity to impose {\em nonlinear} corrections, i.e.,
interactions. 
Therefore, we shall consider a rather general nonlinear low-energy
effective action for the inflaton field 
(interacting quantum fields in classical curved space-times)
and calculate the corrections due to new physics at ultra-high
energies in the anisotropies of the cosmic microwave background.
For simplicity, we assume that the Planck scale is the characteristic
scale for new physics from now on 
(but our results can easily be generalized).

{\em The Model}\quad
%
During slow-roll inflation, the space-time can be approximated by the 
de~Sitter metric ($\hbar=c=1$)
\bea
ds^2 
= 
d\tau^2 - e^{2\Hu\tau} d\BF{r}^2 
= 
a^2(t)[dt^2 - d\BF{r}^2] 
\,,
\ea
with the proper time $\tau$ and the conformal time $t$, respectively,
and the Hubble constant $\Hu$ governing the dynamics of the scale
parameter $a(t)$. 
Assuming locality and covariance 
(non-covariant terms will be discussed later), the most general 
low-energy effective action for a scalar field can be represented in a
gradient expansion and reads
\bea
\mathcal L
=
A(\varphi)+(\partial\varphi)^2B(\varphi)+(\partial\varphi)^4C(\varphi)
+\ord[(\partial\varphi)^6]
\,,
\ea
with $(\partial\varphi)^2=(\partial_\mu\varphi)(\partial^\mu\varphi)$
and arbitrary coefficient functions $A(\varphi)$, $B(\varphi)$, and
$C(\varphi)$. 
Assuming stability $B(\varphi)>0$ and adopting the usual field
re-definition ${d\phi=\sqrt{2B(\varphi)}\,d\varphi}$, we get
\cite{box} 
\bea\label{2}
\mathcal L
=
\frac12\left[(\partial\phi)^2-V(\phi)+(\partial\phi)^4W(\phi)\right]
+\ord[(\partial\phi)^6]
\,.
\label{L}
\ea
The first two terms correspond to the usual scalar field theory whereas 
the third contribution of this low-energy effective action is a remnant 
of new physics (e.g., the interaction with massive gravitons) at 
ultra-high energies.
Identifying the scalar field $\phi$ with the inflaton field, we assume the 
potential for chaotic inflation $V(\phi)=m^2\phi^2$, but our main results 
can easily be generalized to other potentials (see discussion below).
Now we split the inflaton field $\phi$ into a classical background field 
$\phi_0$ plus its quantum fluctuations 
$\phi(t,\BF{r})=\phi_0(t)+\Phi(t,\BF{r})$.
In the usual slow-roll approximation $m \ll \Hu$, the slowly decaying
classical background solution reads 
$\phi_0(\tau) \propto \exp( -m^2 \tau / 3\Hu )$, and the 
Lagrangian for the fluctuations is given by 
\bea
L
&=&
\int d^3r\left\{
\frac{a^2}{2}\left[\dot\Phi^2-(\na\Phi)^2\right]
-ag\left[\dot\Phi^3-\dot\Phi(\na\Phi)^2\right]
\right\}
\nn
&&
+\ord[(\partial\phi)^6]
+\ord[\Phi^4]
+\ord[g^2]
\,,
\label{L_gen}
\ea
with 
$g=2W(\phi_0)\partial\phi_0/\partial\tau\approx-2\phi_0W(\phi_0)m^2/(3\Hu)$ 
denoting a roughly constant effective coupling strength.
Note that the interaction term $(\partial\phi)^4$ in the
Lagrangian~(\ref{L}) also implies small corrections such as
$\dot\Phi^2$ and thus a small re-definition of the propagation 
speed~$c_{\rm eff}$, the mass~$m$, and the amplitude of the inflaton
field $\Phi$ etc.  
As usual during inflation, the mass term $m^2\Phi^2$ can be neglected for 
the fluctuations $\Phi$ due to $m\ll\Hu$ 
(terms like $\lambda\Phi^3$ will be discussed below).
Since we are mainly interested in the three-point function, we omitted
the higher-order terms $\ord[\Phi^4]$, but the associated four-point
contribution can be calculated via the same formalism 
(see discussion below). 
For the sake of brevity, we omit stating the higher-order corrections 
$\ord[(\partial\phi)^6]+\ord[\Phi^4]+\ord[g^2]$ from now on. 

{\em Hamiltonian}\quad
%
After a Legendre transformation, we obtain the low-energy effective 
Hamiltonian density 
\bea
\mathcal H
=
\frac{1}{2}\left[\frac{\Pi^2}{a^2}+a^2(\na\Phi)^2\right]
+g\left[\frac{\Pi^3}{a^5}-\frac{\Pi}{a}(\na\Phi)^2\right]
\,.
\label{H}
\ea
The explicite time-dependence of this Hamiltonian can partly be removed 
via the (time-dependent) canonical transformation $\Pi\to a\Pi$ and 
$\Phi\to\Phi/a$ generated by the squeezing operator $S$, giving
rise to an additional term to the Hamiltonian: 
In the squeezed representation with the Schr\"odinger equation
$i\,d(S\ket{\Psi})/dt = H_S (S \ket{\Psi})$, the Hamilton density
reads
\bea
\mathcal H_S
&=&
\frac{1}{2}\left[\Pi^2+(\na\Phi)^2\right]
-\frac{a\Hu}{2}\left[\Phi\Pi+\Pi\Phi\right]
\nn
&&
+\frac{g}{a^2}\left[\Pi^3-\Pi(\na\Phi)^2\right]
=
\mathcal H_0+\mathcal H_{\Hu}+\mathcal H_g
\,.
\label{H2}
\ea
This form allows for an intuitive physical interpretation:
Due to the cosmic expansion $a=e^{\Hu\tau}$, the modes are being 
continuously stretched and hence the effects of the interactions 
$\mathcal H_g\propto1/a^2$ decrease with time whereas the influence of 
the expansion $\mathcal H_{\Hu}\propto a$ increases.  

{\em Initial state}\quad
%
Since the interactions are strong at early times, the derivation of
the correct initial state requires the treatment of strongly
interacting quantum fields in curved space-times, which is a
nontrivial task, cf.~\cite{birrell}. 
Let us consider the evolution of the modes in more detail:
Due to spatial homogeneity, the co-moving wavenumber $\BF{k}$
corresponding to a spatial dependence of $e^{i\BF{k}\cdot\BF{r}}$
remains constant -- but the physical wavenumber $\BF{k}/a$ decreases
(stretching). 
Hence modes with different $k$ leave the Planck scale at different
times, i.e., when $k/a=\ord[M_{\rm Pl}]$.
Assuming that the rate of the cosmic expansion given by the Hubble
parameter $\Hu$ is much slower than the internal evolution rate of
these Planckian modes (scale separation $\Hu \ll M_{\rm Pl}$), the
most natural initial state is the adiabatic vacuum state,
cf.~\cite{birrell}.  
During the further expansion, the wavelength increases until 
$k/a \ll M_{\rm Pl}$ holds and the low-energy effective
action~(\ref{L}) becomes valid \cite{assume}. 
In view of the adiabatic theorem, the modes evolve adiabatically and 
hence stay in the adiabatic vacuum as long as the external
time-dependence due to the cosmic expansion is slow compared to the
internal dynamics.
If we now choose an intermediate time $t_0$ such that 
$\Hu \ll k/a(t_0) \ll M_{\rm Pl}$, the effects of both, the 
cosmic expansion $\mathcal H_{\Hu}$ and the interaction $\mathcal H_g$
are small and can be treated as perturbations \cite{assume}. 
As a result, the quantum state (adiabatic vacuum) of these modes at
that time $t_0$ can be derived via the adiabatic expansion
\cite{adiabatic} 
\bea
\label{adiabatic}
\ket{{\rm in}(t_0)}
&=&
N_0\ket{\Psi_0}
-i\sum\limits_{n>0}
\ket{\Psi_n}
\left(
\frac{\bra{\Psi_n}\dot H_S\ket{\Psi_0}}{\Delta E_n^2}
\right.
+
\nn
&&
+
\left.
\frac{i}{\Delta E_n}\frac{d}{dt}
\frac{\bra{\Psi_n}\dot H_S\ket{\Psi_0}}{\Delta E_n^2}
\right)
+\ord[\Hu^3]
\,,
\ea
where $\ket{\Psi_n}$ are the instantaneous eigenvectors of the
Hamiltonian~(\ref{H2}) with appropriate phases and $N_0=1+\ord[\Hu^2]$
is a normalization.  
For $g = 0$, the Hamiltonian $H_0+H_\Hu$ can be diagonalized exactly  
(i.e., to all orders in $\Hu$) via a Bogoliubov transformation
$c_{\BF{k}} = \alpha_{\BF{k}} b_{\BF{k}}
+ \beta_{\BF{k}} b_{-\BF{k}}^\dagger$.
Up to second order in $\Hu$ (the accuracy we are interested in), the
Bogoliubov coefficients $\alpha_k$ and $\beta_k$ read 
$\alpha_{\BF{k}} = 1 + i a \Hu / 2 k$ and
$\beta_{\BF{k}} = i a \Hu/ 2 k - a^2 \Hu^2/4k^2$.
The remaining corrections can be calculated with stationary 
perturbation theory
\bea
\ket{\Psi_n}=\ket{\Psi_n^0}+\sum_{m\neq n}\ket{\Psi_m^0}
\frac{\bra{\Psi_m^0}H_g\ket{\Psi_n^0}}{E_n^0-E_m^0}
+\ord[g^2]
\,,
\ea
where $(H_0+H_{\Hu})\ket{\Psi_n^0}=E_n^0\ket{\Psi_n^0}$ are the
unperturbed eigenstates obtained via the aforementioned Bogoliubov
transformation. 

{\em Final state}\quad
%
The impact of the cosmic expansion $H_{\Hu}$ on the modes increases
with time and their evolution finally becomes non-adiabatic when 
$k/a=\ord[\Hu]$, i.e., when they cross the horizon and freeze.
Therefore, in order to calculate the final state, we switch from the 
Schr\"odinger representation (thereby undoing the canonical
transformation $S$ etc.) to the interaction picture: 
The Heisenberg operators carry full dynamics for the unperturbed 
problem $H_0+H_{\Hu}$ to all orders in $\Hu$ (cf.~\cite{text})
\bea
\Phi
=  
\int\frac{d^3k}{\sqrt{2k(2\pi)^3}}
\left(\frac{1}{a} + i \frac{\Hu}{k} \right)
e^{i \BF{k}\cdot\BF{r} - i k/(\Hu a)}
a_{\BF{k}}
+ 
{\rm H.c.}
\ea
Note that $a_{\BF{k}}$ annihilates the adiabatic vacuum
(\ref{adiabatic}) for vanishing $g$, which can be used as a
consistency check. 
The remaining small interaction Hamiltonian $H_g$ acts on the
quantum state, which enables us to calculate the final state 
via time-dependent perturbation theory 
\bea
\ket{{\rm out}(t)}
=
\left(
1-i\int\limits_{t_0}^{t}dt'\,H_g(t)
\right)
\ket{{\rm in}(t_0)}
+\ord[g^2]
\,.
\ea
As another test of the consistency of the used formalism, one may
check that the resulting final state $\ket{{\rm out}(t)}$ is indeed 
independent of $t_0$ as it should be. 
Since we are mainly interested on modes which freeze well before the
end of inflation, we may approximately extend the integration to 
infinite proper time $\tau=\infty$ which corresponds to vanishing 
conformal time $t=0$.

{\em Three-point function}\quad
%
With the final state obtained after performing the integration,  
one can calculate the three-point function of the frozen fluctuations 
and obtains (after some algebra) the following spectrum
\wbea
\label{3}
\langle\Phi(\BF{r}_1)\Phi(\BF{r}_2)\Phi(\BF{r}_2)\rangle
&=&
-g\Hu^5
\int\frac{d^3k_1\,d^3k_2\,d^3k_3}{(2\pi)^6}
\,
\exp\{i(\BF{k}_1\cdot\BF{r}_1+\BF{k}_2\cdot\BF{r}_2+\BF{k}_3\cdot\BF{r}_3)\}
\,
\frac{\delta^3(\BF{k}_1+\BF{k}_2+\BF{k}_3)}{k_1k_2k_3}
\times
\nn
&&
\times
\left(
\frac{3}{k_+^3}
-
\left\{
\frac{\BF{e}_2\cdot\BF{e}_3}{k_+^3}
+\frac{1}{k_+}\frac{\BF{e}_2\cdot\BF{e}_3}{2k_2k_3}
+\frac{\BF{e}_2\cdot\BF{e}_3}{2k_+^2}\left(\frac{1}{k_2}+\frac{1}{k_3}\right)
+{\rm perm.}
\right\}
\right)
\left(1+\ord[\Hu]+\ord[g]\right)
\,,
\wea
with the abbreviations $k_+=k_1+k_2+k_3$ and $\BF{e}_j=\BF{k}_j/k_j$.
The remaining permutations (denoted by ``perm.'', i.e.,  
$k_1 \to k_2$, $k_2 \to k_3$, and $k_3 \to k_1$ etc.) 
have to be included as well. 
The spectrum is scale invariant as one would expect 
since the sequence: leaving of Planck scale and stretching $\to$
horizon crossing and freezing occurs on all scales 
(just at different times).
The first contribution $3/k_+^3$ on the second line results from the
$\Pi^3$-term whereas the remaining ones appear because of the
$\Pi(\na\Phi)^2$-term in the interaction Hamiltonian $H_g$. 

An estimate of the relative size of this non-Gaussian contribution can
be obtained by comparison to the (Gaussian) two-point function
\bea
\label{relative}
\frac{\langle\Phi\Phi\Phi\rangle}{\sqrt{\langle\Phi\Phi\rangle^{3}}}
=\ord[g\Hu^2]
=\ord[\phi_0W(\phi_0)m^2\Hu]
\ll1
\,.
\ea
Note that $m$ can be eliminated with the classical equations of motion
and the Friedman equation, i.e.,  
$g\Hu^2=\ord[\phi_0W(\phi_0)m^2\Hu]
=\ord[W(\phi_0)\Hu^3 M_\mathrm{Pl}^2/\phi_0]$.
Usually the typical scale of new physics such as quantum gravity is
expected to be the Planck scale.
Therefore, one would naturally assume that the coupling behaves
according to $W\sim1/M_{\rm Pl}^4$ if $W$ did not depend on $\phi_0$. 
In this case, the relative size of the effect would be suppressed with 
$\Hu^3/M_\mathrm{Pl}^3$ and thus probably hard to measure.

However, one should also bear in mind that the usual picture of
inflation involves a large value of the inflaton field 
$\phi_0\gg M_{\rm Pl}$ and hence $W(\phi_0)$ could be much bigger than 
$1/M_{\rm Pl}^4$. 
Similarly, the scale for new physics might be below the Planck scale
or the inflaton potential $V(\phi)$ could differ from $m^2\phi^2$. 
In all these cases, the effect could be far stronger than   
$\Hu^3/M_\mathrm{Pl}^3$.
On the other hand, the employed formalism only works if $g\Hu^2\ll1$
holds, i.e., if the relative size of the effect~(\ref{relative}) 
is small compared to one 
(which is consistent with the observations \cite{wmap}). 

{\em Generalizations}\quad
%
Let us discuss the impact of momentum-independent interaction terms
such as $\lambda\Phi^3$ 
[in contrast to contributions like $(\partial\Phi)^4$]
which occur if the potential $V$ is not just given by the mass
term $V(\phi)\neq m^2\phi^2$. 
In contrast to $H_g$, these momentum-independent interaction terms
become important at late times.
Hence the three-point spectrum generated by the $\lambda\Phi^3$-term 
can be derived from time-dependent perturbation theory and behaves as 
$k_+/(k_1 k_2 k_3)^3$.
This spectrum is dominated by large length scales and thus not scale 
invariant -- which should make it easy to distinguish between the 
effects of the momentum-independent interaction $\lambda\Phi^3$ and
those generated by high-energy (Planckian) interactions such as 
the spectrum~(\ref{3}).

The derivation of the four-point function can be accomplished by means
of the same formalism -- but necessitates higher orders in the
adiabatic expansion etc.
The relative size of the non-Gaussian contribution (compared to the
Gaussian part of the four-point function, for example) can be 
estimated to $\ord[W(\phi_0)\Hu^4]$.
Hence the comparison of the relative sizes of the non-Gaussian
contributions of the three-point and the four-point functions,
respectively, yields the unknown ratio $\Hu\phi_0/M_\mathrm{Pl}^2$.

{\em Lorentz invariance}\quad
%
So far, we assumed that the Planckian interactions respect the local
Lorentz invariance and hence started from a covariant action (\ref{L}).
However, it may well be that the usual local Lorentz invariance is
just a low-energy symmetry and broken at the Planck scale.
In this case, many more terms can occur in the low-energy effective
action -- even if we still demand spatial homogeneity and isotropy
according to the cosmological principle:
First of all, bilinear terms such as $\ddot\Phi^2$ and $(\na^2\Phi)^2$
exactly correspond to a change in the dispersion relation mentioned in
the introduction.
Assuming adiabaticity (which may not by valid for all dispersion
relations), we may treat those corrections with the presented
formalism and it turns out that they basically generate a global shift
of the two-point spectrum to lowest order, which is probably  
hard to measure. 

The next terms occurring in an expansion into powers of $\Phi$ and the
numbers of derivatives are exactly the terms $\Pi^3$ and
$\Pi(\na\Phi)^2$ discussed before \cite{multiple}.
However, without Lorentz invariance, the pre-factors of these two
terms are no longer necessarily equal.
Therefore, measuring and comparing the relative strength of the two
contributions in the above three-point spectrum~(\ref{3}) provides a
possibility to detect violations of the Lorentz invariance at the
Planck scale.

{\em Summary}\quad
%
Assuming locality and covariance, we derived the most general
low-energy effective action of the scalar inflaton field 
(including possible remnants from Planckian interactions) 
via a gradient expansion. 
Based on this ansatz, a formalism for the treatment of this (strongly)
interacting quantum field theory in a curved space-time and the
derivation of the quantum state via the adiabaticity assumption was
presented.  
This enabled us to predict the induced spectrum of the three-point
function -- which is quite robust against uncertainties in cosmic
evolutions and can clearly be distinguished from other effects such as
a $\lambda\Phi^3$-term -- making it a promising observable.  
This concrete prediction of the explicite spectrum facilitates a
refined search in the present/future WMAP and PLANCK data with good
statistics and hence high sensitivity. 
Abandoning covariance, it might even be possible to detect violations
of the Lorentz invariance at the Planck scale. 

\acknowledgments
{\em Acknowledgments}\quad
%
This work was supported by the Emmy-Noether Programme of the
German Research Foundation (DFG) under grant No.~SCHU 1557/1-1,2.
Further support by the COSLAB programme of the ESF and 
the Pacific Institute of Theoretical Physics is also gratefully
acknowledged. 



\begin{thebibliography}{499}
\bibitem{box}
To first order (in $W$ or $g$), which is the accuracy we are
interested in, one can eliminate terms containing $\Box\phi$ 
(which would complicate the Hamiltonian analysis) after an integration 
by parts as well as using the equation of motion for $\phi$.
%
The same applies to contributions of the type  
$\phi^2(\partial_\mu\partial_\nu\phi)(\partial^\mu\partial^\nu\phi)$
etc.
\bibitem{assume}
Here we assume $g=\ord(1/\sqrt{M_{\rm Pl}})$, but it would be
sufficient to demand $g\Hu^2\ll1$ and to replace $M_{\rm Pl}$ by 
$1/\sqrt{g}$, i.e., $\Hu \ll k/a(t_0) \ll 1/\sqrt{g}$.
\bibitem{adiabatic}
This expansion can be derived via the decomposition 
$\ket{\Psi(t)}=\sum_n a_n(t) \exp\{-i\int dt' E_n(t')\} \ket{\Psi_n(t)}$
of the quantum state $\ket{\Psi(t)}$ into the instantaneous
eigenstates $\ket{\Psi_n(t)}$ and insertion of this ansatz into the
Schr\"odinger equation $id\ket{\Psi(t)}/dt=H(t)\ket{\Psi(t)}$.
%
After an integration by parts, the coefficients $a_n(t)$ can be
expressed in terms of the matrix elements 
$\bra{\Psi_n}\dot H\ket{\Psi_m}$ divided by the energy gap squared  
$(E_n-E_m)^2$.
%
For adiabatic evolution, this ratio must be small which facilitates an 
expansion into powers of $\Hu$.
\bibitem{multiple}
In case of multiple inflaton fields, many more combinations are
possible -- in both cases: with and without Lorentz invariance.
%
Furthermore, we assume that the scale factors $a(t)$ occur with the
same power as in the Lorentz invariant situation -- which is not
strictly necessary without demanding covariance.
\end{thebibliography}
\end{document}